\documentclass[aps,twocolumn,notitlepage,pra,superscriptaddress,floatfix]{revtex4-2}
\usepackage{academicons}
\usepackage{graphicx}
\usepackage{braket}
\usepackage{amsmath,amsfonts,amssymb,amstext,amscd,amsthm,bbold}
\usepackage{comment,float}
\newcommand{\tr}{\textcolor{red}}

\usepackage{xcolor}
\definecolor{burgundy}{RGB}{128,0,32}
\usepackage[colorlinks=true,linkcolor=blue,urlcolor=magenta,citecolor=magenta]{hyperref}
\hypersetup{
    colorlinks=true,
    linkcolor=blue,
    filecolor=red,      
    urlcolor=burgundy,citecolor=purple
}
\usepackage[normalem]{ulem}
\usepackage{multibib}
\newcites{book,misc}{{Books},{Others}}
\def\beq{\begin{eqnarray}}\def\eeq{\end{eqnarray}}
\def\be{\begin{equation}}\def\ee{\end{equation}}
\def\a{\alpha}

\def\d{\delta}

\def\g{\gamma}

\def\l{\lambda}
\def\m{\mu}

\def\r{\rho}
\def\s{\sigma}

\def\la{\langle}
\def\ra{\rangle}

\def\ua{\uparrow}
\def\da{\downarrow}

\def\tr{{\rm tr~}}

\usepackage{bm}
\usepackage{bm}

\usepackage{xcolor}

\definecolor{orcidlogocol}{HTML}{A6CE39}
\usepackage{silence}
\usepackage{orcidlink}
\definecolor{mycolor}{HTML}{29BF12}
\usepackage[normalem]{ulem}

\renewcommand{\selectlanguage}[1]{}

\usepackage{titlesec}
\titleformat{\section}
{\sffamily\bfseries} 
{\thesection} 
{1em} 
{} 

\titleformat{\subsection}
{\sffamily\bfseries} 
{\thesubsection} 
{1em} 
{} 

\titlespacing\section{0pt}{12pt plus 4pt minus 2pt}{2pt plus 2pt minus 2pt}
\titlespacing\subsection{0pt}{12pt plus 4pt minus 2pt}{2pt plus 2pt minus 2pt}
\titlespacing\subsubsection{0pt}{12pt plus 4pt minus 2pt}{2pt plus 2pt minus 2pt}

\begin{document}
	\title{Complexity for one-dimensional discrete-time quantum walk circuits}
	
	 \author{Aranya Bhattacharya$^{\orcidlink{0000-0002-1882-4177}}$}
	 \affiliation{Institute of Physics, Jagiellonian University, Łojasiewicza 11, 30-348 Kraków, Poland} \affiliation{Center for High Energy Physics,
	Indian Institute of Science, C.V. Raman Avenue, Bangalore 560012, India}
	 
	\author{Himanshu Sahu$^{\orcidlink{0000-0002-9522-6592}}$}
	\email{himanshusah1@iisc.ac.in}
	\affiliation{Department of Instrumentation \& Applied Physics, Indian Institute of Sciences, C.V. Raman Avenue, Bengaluru 560012, India}
	\affiliation{Center for High Energy Physics,
		Indian Institute of Science, C.V. Raman Avenue, Bangalore 560012, India}
    
    \author{Ahmadullah Zahed$^{\orcidlink{0000-0003-1252-8171}}$}
	 \affiliation{Center for High Energy Physics,
	Indian Institute of Science, C.V. Raman Avenue, Bangalore 560012, India}

	\author{Kallol Sen$^{\orcidlink{0000-0002-5540-0285}}$}
	\email{kallolmax@gmail.com}
	\affiliation{ICTP-South American Instutute of Fundamental Research, IFT-UNESP (1º andar), Rua Dr. Bento Teobaldo Ferraz 271, Bloco 2 - Barra Funda 01140-070 São Paulo, SP Brazil}

	\begin{abstract}
		We compute the complexity for the mixed state density operator derived from a one-dimensional discrete-time quantum walk (DTQW). The complexity is computed using a two-qubit quantum circuit obtained from canonically purifying the mixed state. We demonstrate that the Nielson complexity for the unitary evolution oscillates around a mean circuit depth of $k$. Further, the complexity of the step-wise evolution operator grows cumulatively and linearly with the steps. From a quantum circuit perspective, this implies a succession of circuits of (near) constant depth to be applied to reach the final state. 
		

	\end{abstract}
	\maketitle
	
	\section{Introduction}
	
	Nielsen's complexity (NC)\cite{Nielsen_2006,dowling2006geometry,jefferson_circuit_2017, Hackl:2018ptj, Camargo_2019, Balasubramanian:2019wgd, Bhattacharya:2022wlp}, which is conjectured to quantify the optimal number of quantum gates needed (with a pre-decided set of elementary gateset) to construct a target state starting from a reference state, has been of particular interest to the high energy physics community in recent years. It was also suggested in\,\cite{jefferson_circuit_2017} to be related to circuit depths since  circuit depth in an actual quantum circuit is also a measure (although not necessarily optimal) of the number of gates needed to implement a certain task. However, the exact connection of the NC measure with the number of gates in quantum circuits is far from being fully understood. The reason is the ambiguity of precisely mapping the complexity measure to the quantum circuit picture. A better understanding of a possible link between circuit depth and Nielsen's complexity proposal could provide an analytical handle on the practical circuit building using quantum gates and ask whether the circuit in question is optimal. From the reverse point of view, it is only logical to bring the analytically well-defined notion of circuit complexity proposal closer to actual circuits in quantum simulations. Otherwise, relating the mathematically computed NC to something physically meaningful becomes hard. This work explores this question from the discrete-time quantum walk (DTQW) perspective.\\
	
	 Quantum walks, a quantum mechanical analog to classical random walks\,\cite{venegas-andraca_quantum_2012,venegas-andraca_quantum_2012,ambainis2004quantum}, provide a powerful framework in quantum computing for algorithms like quantum search and optimization\,\cite{PhysRevA.70.022314,Childs_2003,camposQuantumMetropolisSolver2023,Slate2021quantumwalkbased}. They are also proven successful frameworks for modeling quantum systems, such as simulating Dirac equations\,\cite{doi:10.1126/science.273.5278.1073,huertaaldereteQuantumWalksDirac2020a,oka_breakdown_2005} as well as modeling biological processes\,\cite{engelEvidenceWavelikeEnergy2007,Rebentrost_2009}. In a discrete-time quantum walk, a quantum particle or a qubit is allowed to move along a discrete graph or lattice in discrete-time steps. At each discrete-time step, the particle state undergoes a unitary operation that consists of two operators, namely the coin and the shift operators. The coin operator operates as a rotation within the qubit space, while the shift operator serves to translate the particle to another vertex within the lattice. DTQWs have emerged as pivotal tools in the realm of quantum information processing, finding widespread applications in the development of diverse quantum algorithms\,\cite{shenvi_quantum_2003,portugal_quantum_2013,ambainis2004coins}, modeling of quantum systems\,\cite{chandrashekar_two-component_2013,mallick_dirac_2016,mallick_simulating_2019,PhysRevA.73.054302,mallick_neutrino_2017,molfetta_quantum_2016,Sahu:2023csa}, and extensively investigated across various settings\,\cite{Sen:2023zyx, Zahed:2023thw, Omanakuttan:2021aa}. DTQWs have been successfully implemented in experimental settings using lattice-based quantum systems, wherein the position space is mapped onto discrete lattice sites\,\cite{flurin_observing_2017,ramasesh_direct_2017,peruzzo_quantum_2010,tamura_quantum_2020,karski_quantum_2009,broome_discrete_2010,perets_realization_2008,zahringer_realization_2010,schreiber_photons_2010}. Additionally, these quantum walks have been realized through circuit-based quantum processors, which allow the computation of the complexity of the circuit\,\cite{qiang_efficient_2016,ryan_experimental_2005,huerta_alderete_quantum_2020}.\\

  However, we are not concerned directly with the circuit implementation of the DTQW for this work. Instead, we focus on computing the complexity of the qubit state (or coin state) associated with the quantum particle. To this end, we trace out the position space and focus on the reduced density matrix associated with the qubit (see Eqs.~\eqref{DTQWF} and \eqref{mixeddensity}). The complexity of the single qubit mixed state is then computed by canonically purifying the reduced density matrix, after which it becomes a two qubit pure state. This purification process can be implemented by a corresponding two-qubit circuit and that will be our main interest in this paper.

\begin{figure}
		\centering
		\includegraphics[scale = 0.25]{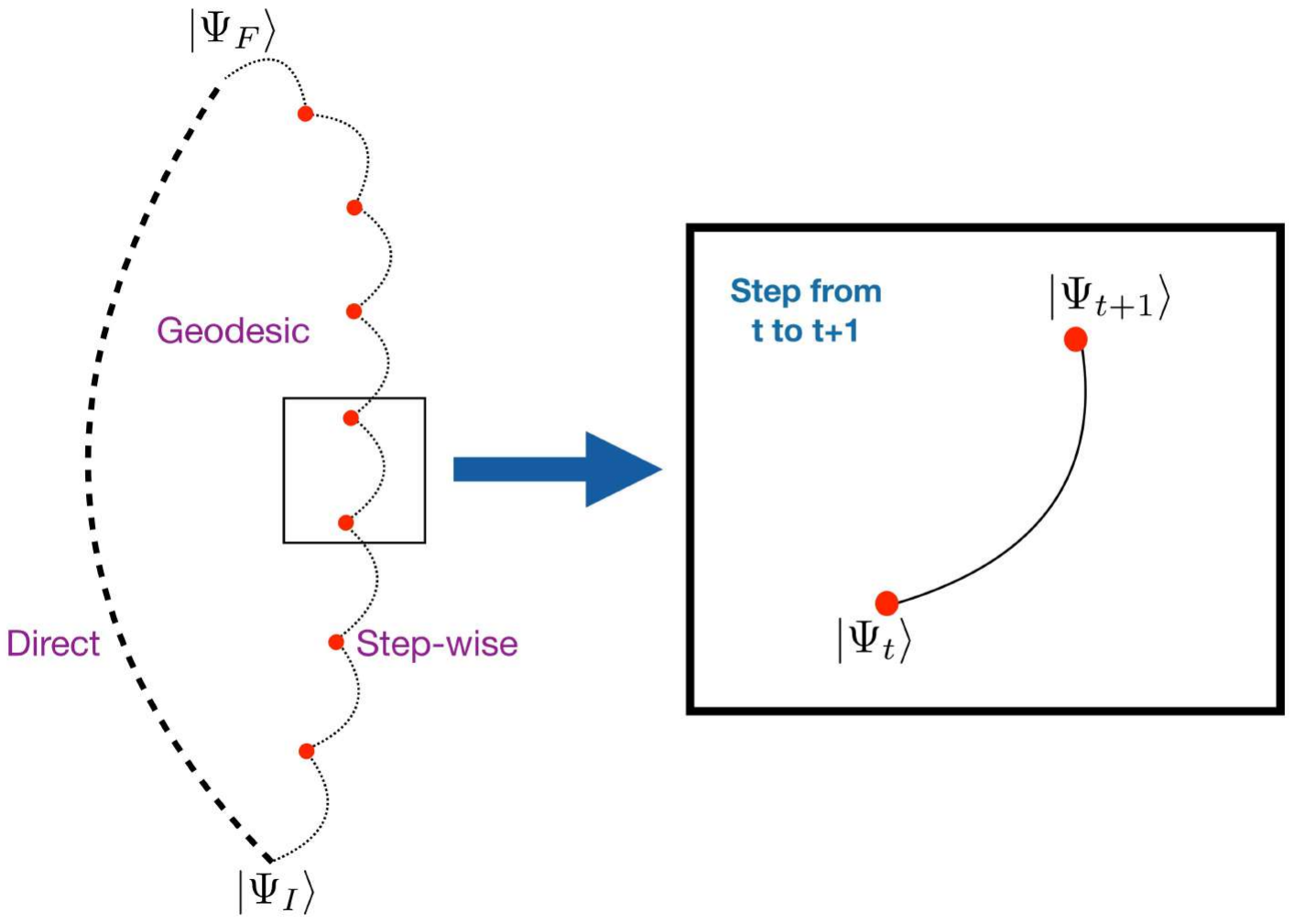}
		\caption{Comparing the direct and step-wise evolution from an initial state $|\Psi_I\ra$ to $|\Psi_F\ra$.}
		\label{fig:geodesic}
	\end{figure}

 At this point, it is worth pointing out that NC measures the length of the minimal geodesic in the space of ``response functions"\footnote{These quantify the number of times a particular gate is used in a specific time.} from the initial to the final state. However, a DTQW is implemented step-wise, and the information about the step-wise evolution is lost if we naively compute the geodesic length (direct complexity) connecting the initial and final steps. We refer the reader to the schematic diagram in Fig.~\ref{fig:geodesic} for now. This difference between the direct and stepwise complexities will be crucial in associating the complexity with the circuit depth of practically realizable quantum circuits. We will get back to the details of these two definitions later in the paper.\\

	\textbf{Results:} The main results of our work are comparing the two approaches for determining the complexity for the $1D$ DTQW using step-wise and direct evolution. This comparison is done for $k=1,2,3$ local operators in Sec.\,\ref{twodef}. Here the $k$-local notion comes from the number of Majorana fermionic operators considered in constructing the generators. We construct explicit solutions of geodesic trajectories in the space of the unitaries 
    as a function of the affine parameter by solving the geodesic equation\,\eqref{euler_arnold} for the $k$-local cases. These solutions are useful in constructing the target unitary operator, albeit their functional forms are irrelevant to the complexity computation.
	
	The complexity for the direct evolution follows an oscillatory pattern around a mean depth, which varies with the locality of operators. This indicates that the walk is truly random, and the complexity for different states at different times compared to a single reference state does not show any correlated growth. The step-wise evolution, on the other hand, is linear in steps,
	where the slope is a function of the coin angle.
	The step-wise complexity can also indeed be mapped to a quantum circuit of constant circuit depth and composed of universal quantum gates. This study, therefore, enables a comparison between the Nielsen complexity and the quantum circuit depth. Although they do not correspond to the same quantity mathematically, the similarities in scaling suggest an approximate relation between them. \\
	\\
	The remainder of the work is organized as follows. In Sec.~\ref{1DTQW}, we give an account of the $1d$ DTQW using the $\text{SU}(2)$ coin. In Sec.~\ref{Ut}, we construct the unitary target operator that takes a simple unentangled reference state to our desired canonically purified target state.
In Sec.~\ref{complexity}, we construct the explicit form of the unitary operator by looking at the explicit $k-$local solutions to the geodesic equation. Specifically, in Sec.~\ref{twodef}, we compare the two distinct methods to compute the complexity using the direct evolution and step-wise evolution of the walk. In Sec.~\ref{circuit}, we give an elementary account of the quantum circuit that represents the step-wise unitary operator $U_\text{step}$, using elementary one and two-qubit gates $U(\theta,\phi,\lambda)$ and CNOT gates. We end the work with discussions in Sec.~\ref{discussions} on the questions we have left unanswered in the work and wish to complete them in future follow-ups.

	\section{\texorpdfstring{$1D$}{Lg} Discrete Time Quantum Walk}\label{1DTQW}

\begin{figure}
		\centering
		\includegraphics[scale=0.5]{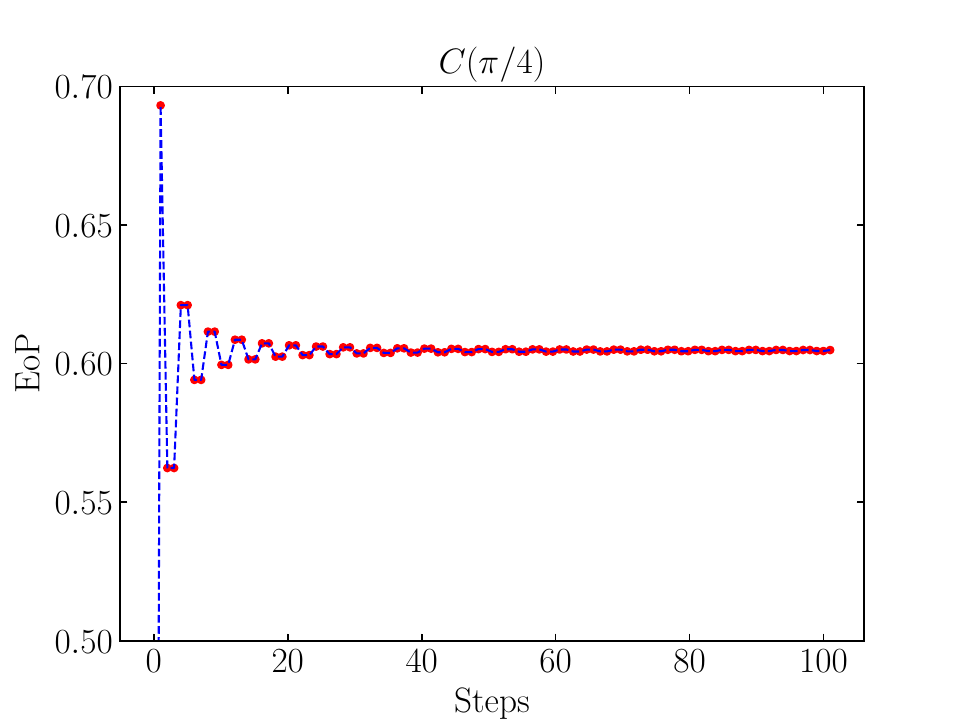}
		\caption{Entanglement of purification (EoP) with steps for canonical purification.}
		\label{fig:EOP}
	\end{figure}

	The $1D$ discrete-time quantum walk (DTQW) we study is given by the unitary evolution of a quantum state on a line. The evolution is governed by the operator,
	\be
	U=S\cdot\left(C(\theta)\otimes\mathbb{I}_N\right)\,,
	\ee
	where $S$ is the shift operator,
	\be
	S=\sum_x \left(\lvert \ua,x+1\ra\la\ua,x\rvert+\lvert\da,x-1\ra\la\da,x\rvert\right)\,,
	\ee
	where $\lvert \ua\ra\,,\lvert\da\ra$ are the directions of motion (to the left or right) of a particular node $x\in N$ where $N$ is the number of nodes. The coin operator $C$ given by,
	\be
	C(\theta)=\sum_{i,j\in \ua,\da}c_{ij}(\theta)|i\ra\la j|\,.
	\ee
	controls the weights of motion in particular directions. For our case, we consider a subset of the $\text{SU}(2)$ coin operator,
	\be
	C(\theta)=\left(\begin{array}{cc} \cos\theta & \sin\theta\\ -\sin\theta&\cos\theta\end{array}\right)\,,
	\ee
	where we have set the phases to zero without loss of generality. The unitary operator lives in $\mathcal{H}_2\otimes\mathcal{H}_N$ which is a $(4N+2)\times(4N+2)$ dimensional space. The initial state is chosen to be positioned at the origin with an equal superposition of the coin states,
	\be
	|\Psi(0)\ra=\frac{\lvert\ua\ra+i\lvert\da\ra}{\sqrt{2}}\otimes |0\ra\,.
	\ee
	After $t$ steps of evolution,
	\be \label{DTQWF}
	|\Psi(t)\ra=U^t|\Psi_0\ra\,,
	\ee
	can be written as a general superposition of the $\lvert\ua\ra$ and $\lvert\da\ra$ states,
	\be
	|\Psi(t)\ra=\sum_x \left(A_x(t)\lvert\ua,x\ra+B_x(t)\lvert\da,x\ra\right)\,.
	\ee
	The coefficients can be recursively solved from the relations,
	\be
	\begin{split}
			A_x(t)&=\cos\theta A_{x-1}(t-1)+\sin\theta B_{x-1}(t-1)\,, \\ B_x(t)&=-\sin\theta A_{x+1}(t-1)+\cos\theta B_{x+1}(t-1)\,,
	\end{split}
	\ee
	The probability distribution as a function of the time and position index is given by $p_x(t)=\lvert A_x(t)\rvert^2+\lvert B_x(t)\rvert^2$. To proceed, we consider the reduced density matrix, 
	\begin{align}\label{mixeddensity}
		\begin{split}
			&\r(t)=\tr_x|\Psi(t)\ra\la\Psi(t)|=\sum_{i,j\in\ua,\da}\r_{ij}|i\ra\la j|\,, \\
			&\r_{\ua\ua}=\sum_x |A_x(t)|^2\,,\\
			&\r_{\ua\da}=\sum_x A_x(t)B_x(t)^\star\,, \\
			&\r_{\da\ua}=\r_{\ua\da}^\star\,, \r_{\da\da}=1-\r_{\ua\ua}\,.
		\end{split}
	\end{align}
	By construction, $\tr\r(t)=1$, and the resultant is a mixed state density matrix in the coin space $(\mathcal{H}_2)$. We start by canonically purifying the reduced density matrix, which begins at computing the eigenvalues of the matrix $\r(t)$, given by,
	\be
	\l_\pm(t)=\frac{1\pm\sqrt{1-4\det\r}}{2}\,.
	\ee
	and corresponding eigenvectors $|\psi_\pm\ra$. The resultant canonically purified state,
	\be\label{pure_state}
	|\Phi(t)\ra=\sqrt{\l_+(t)}|\psi_+,\psi_+\ra+\sqrt{\l_-(t)}|\psi_-,\psi_-\ra\,,
	\ee
	is a two-qubit state where $|\psi,\psi\ra=|\psi\ra\otimes|\psi\ra$. This is our starting point for the complexity computation. It is motivated by the following principle. Since complexity computation is known for pure states, the corresponding evaluation for mixed states entails an additional intermediate step of purifying the mixed state to a pure state at the cost of dimensional oxidation from $\mathcal{H}^{2^n}\to \mathcal{H}^{2^{2n}}$ space. The corresponding entanglement of purification as a function of time is,
	\be
	\text{EoP}(t)=-\text{tr}\left(\rho_{pr}\ln\rho_{pr}\right)\,,\ 
	\ee
	$\text{where}\ \r_{pr}=\text{tr}_2\left(|\Phi(t)\ra\la\Phi(t)|\right)\,.$ The density matrix $\r_{pr}$ is the reduced density matrix from the purified state where $\text{tr}_2$ implies the partial trace of the second qubit. The functional dependence of entanglement of purification on steps is given in Fig.~\ref{fig:EOP}. It follows essentially the same behavior as the entanglement for the quantum walk.
	
	To conclude the section, we would like to comment on the continuum limit of the walk and its implications on purification. As was established in\,\cite{DiMolfetta:2019kgj, chandrashekar_two-component_2013}, the continuum limit of the one-dimensional walk is given by the Dirac-Hamiltonian for a single free fermion, 
	\be
	\begin{split}
	H(p) &=-i p\left(
	\begin{array}{cc}
		\cos \theta  & \sin \theta  \\
		\sin \theta  & -\cos \theta  \\
	\end{array}
	\right) \\ 
	& \qquad \qquad \qquad \qquad  +\left(
	\begin{array}{cc}
		0 & -i \sin \theta  \\
		i \sin \theta  & 0 \\
	\end{array}
	\right)
	\end{split}
	\ee
	This Hamiltonian characterizes a pure fermionic state. To put this loosely, we can construct a two-particle state (also a pure state) by,
	\be\label{2phamiltonian}
	H(p_1,p_2)=H(p_1)\otimes\mathbb{I}_2+\mathbb{I}_2\otimes H(p_2)\,.
	\ee

\section{Target Unitary Operator}\label{Ut}

	Our starting point is the construction of a unitary operator $U_\text{target}$ such that,
	\be\label{target_unitary}
	|\Phi(t)\ra=U_\text{target}|\Psi_R\ra\,,
	\ee
	where $|\Phi(t)\ra$ is the target state given in \eqref{pure_state} and $|\Phi_R\ra$ is a reference state chosen to be the most simple two-qubit state $|0\ra\otimes|0\ra$. We call $U_\text{target}$ as the {\it target unitary operator} that converts the reference state into the target state. Since $|\Phi(t)\ra$ is time-dependent, we expect $U_\text{target}$ to be a time-dependent matrix as well. However, the matrix is not fully constrained by \eqref{target_unitary}. We begin by constructing,
	\be
	U_\text{target}=\left(\textbf{u}\,, \textbf{u}_1\,, \textbf{u}_2\,, \textbf{u}_3\right)\,,
	\ee
	where $\textbf{u}_i$ are column vectors of dimension $4\times1$. From \eqref{target_unitary}, we get $\textbf{u}=|\Phi(t)\ra$, satisfying $|u|^2=1$ from normalization of $|\Phi(t)\ra$. The unitarity constraint $UU^\dagger=\mathbb{I}$ enforces,
	\be
	\textbf{u}_i\cdot \textbf{u}_j^\dagger=\d_{ij}\,,
	\ee
	which is incidentally the condition for Gram-Schmidt orthonormalization. To start with, 
	\be
	\textbf{u}_i=\textbf{v}_i-\sum_{j=1}^{i-1}\frac{\la \textbf{u}_j,\textbf{v}_i\ra}{\lVert\textbf{u}_j\rVert^2}\textbf{u}_j\,.
	\ee
	where $\lVert\dots\rVert$ is the norm of the vector. We choose $\textbf{v}_i$ to be random vector, 
	\be
	\textbf{v}_i=\textbf{a}_i+i \textbf{b}_i\,,
	\ee
	where $\textbf{a}_i\,, \textbf{b}_i\in \text{rand}_{(0,1)}$. This approach is identical to considering the vectors $\textbf{u}_i\in \text{SU}(4)$ and then optimizing over the parameters. This optimization is over 15 parameters which coincide with the parameterization of $U_\text{target}\in \text{SU}(4)$. We perform the optimization numerically by sampling over $n$ samples of choices of the initial random vectors $\textbf{v}_i$. The number of samples for this optimization depends on whether the standard deviation of the complexity computed from each sample reaches saturation. We will report on this saturation of standard deviation in the next section after discussing the notion of complexity.  
	
	\section{Complexity}\label{complexity}
 
	Once the unitary target operator $U_\text{target}$ is determined, we proceed to compute the complexity of the operator $C(U_\text{target})$. The idea is simple. To implement the purified state, one needs to construct a quantum circuit. The complexity measure $C$ determines the cost of constructing such a quantum circuit. One could argue that we can as well determine the complexity of the quantum circuit implementing the walk. But the quantum walk gives rise to a mixed state density matrix in the coin space. Hence a refined quantum circuit is needed, which first implements the purification of the mixed state. The refined quantum circuit is a two-qubit circuit, as opposed to the one-qubit circuit for the quantum walk. We begin by constructing a path ordered unitary operator for a two-qubit circuit,
	\be
	U(s)=\mathcal{P}\exp\left(-i\int_0^s ds' V_i(s')T_i\right)\,,
	\ee
	where, $V_i(s)$ measures the response function for the generators $T_i\in \text{SU}(4)$ group. It is unclear, as of yet, how these response functions are connected to the strengths (or, more specifically, numbers) of quantum gates needed to construct an actual quantum circuit. However, at this point, we will refrain from addressing this issue and will come back to this at the end of the work. $\mathcal{P}$ denotes path ordering, which denotes the non-commutativity of quantum gates. The generators $T_i$ are built from Majorana fermionic operators $\g_a$, satisfying,
	\be
	\{\g_a, \g_b\}=2\d_{ab}\,.
	\ee
	The generators $T_i$ are given by,
	\be
	T_i=i^{\binom{q}{2}}\g_1^{b_1}\g_2^{b_2}\g_3^{b_3}\g_4^{b_4}\,,
	\ee
	where $b_i$ are the bitwise representation of the integers representing the generators,
	\be
	1\leq i=2^3b_4+2^2b_3+2^1b_2+2^0b_1\leq 15\,, 
	\ee
	and $q=b_1+b_2+b_3+b_4$. The fermionic generators $\g_a$\,\cite{Kuusela:2019aa} and explicit forms of the generators $T_i$ are given in the Appendix~\ref{generators}. We also define the structure constant and the Cartan killing forms ($h=32$ is the Coxeter number),
	\be
	f_{ij}^k=-\frac{i}{4}\tr T_k[T_i,T_j]\,, \ \ K_{ij}=-\frac{1}{h}f_{il}^m f_{jm}^l\,.
	\ee
	Associated with quantum gates is a notion of locality. We import the concept of $k-$locality of operators by associating a cost function $c_i$ such that $c_i=1$ whenever $T_i$ is built from $k$ or fewer $\g_a$ otherwise $c_i=1+\m$ with $\m\gg1$. The cost (or penalty) functions denote the cost of constructing the equivalent quantum gates (so-called `easy' or `hard' gates). With the above definitions, we construct the bi-linear invariant metric on the space of $\text{SU}(4)$ operators,
	\be
	G_{ij}=\frac{c_i+c_j}{2}K_{ij}\,.
	\ee
	With the choice of normalization $K_{ij}=\d_{ij}$ and hence $G_{ij}=c_i\d_{ij}$. The quadratic cost function that defines the complexity is given by,
	\be
	C(U)=\text{min}\int_0^1 ds\sqrt{G_{ij}V^i(s)V^j(s)}\,,
	\ee
	where the functions $V^i(s)$ satisfy the Euler-Arnold geodesic equations,
	\be\label{euler_arnold}
	G_{ij}\frac{dV^j(s)}{ds}=f_{ik}^pG_{pl}V^k(s)V^l(s)\,.
	\ee
	The minimization is over all geodesics leading the affine path from $s=0$ to $s=1$. The minimization takes the geodesic solution to the Euler-Arnold equation \eqref{euler_arnold}. 
	Note that the explicit solutions are relevant for the construction of the unitary matrix, but in so far as the complexity is concerned, only the sums of squares of the functions are important. However, depending on the solutions, the sum of squares of the functions form simple subsets which are constants and independent of $s$. In this sense,
	\be
	C(U)=\sqrt{A^T A+B^T B+\cdots}\,,
	\ee
	where $A, B$, and so on are the subsets. These solutions can be obtained by matching,
	\be
	U(s=1)=U_\text{target} \Rightarrow\ V_i(s=1) T_i=i\ln [U_\text{target}]
	\ee
	In the next few sections, we will solve \eqref{euler_arnold} explicitly for $k=1,2,3$ local cases and construct the complexity explicitly. 
	
	\subsection{\texorpdfstring{$k=1$}{Lg}}
 
	For this case, we have the constants of motion $\left(V_i(s)=v_i\right)$ in the subset,
	\be
	\mathcal{B}=\left(v_5\,,v_6\,,v_7\,,v_8\,,v_9\,,v_{10}\right)\,.
	\ee
	The remaining equations are of the form,
	\be
	\begin{split}
		\frac{d\mathcal{A}_1(s)}{ds}+2\m\mathcal{M}_1\mathcal{A}_1(s)&=0\,,\\ \frac{d\mathcal{A}_2(s)}{ds}+\frac{2\m}{1+\m}\mathcal{M}_2(s)\mathcal{A}_2(s)&=0\,,
	\end{split}
	\ee
	where $\mathcal{A}_1(s)=\left(V_1(s)\,,V_2(s)\,, V_3(s)\,,V_4(s)\right)$, $\mathcal{A}_2(s)=\left(V_{11}(s)\,,V_{12}(s)\,, V_{13}(s)\,,V_{14}(s)\,,V_{15}(s)\right)$ and 
\vspace{1em}
\begin{widetext}
	\begin{equation}
		\mathcal{M}_1(s)=\left(
\begin{array}{cccc}
	0 & v_5 & v_6 & v_8 \\
	-v_5 & 0 & v_7 & v_9 \\
	-v_6 & -v_7 & 0 & v_{10} \\
	-v_8 & -v_9 & -v_{10} & 0 \\
\end{array}
\right)\,, \ \ \  	 \mathcal{M}_2(s)=\left(
\begin{array}{ccccc}
	0 & 0 & 0 & 0 & -V_4(s) \\
	0 & 0 & 0 & 0 & V_3(s) \\
	0 & 0 & 0 & 0 & -V_2(s) \\
	0 & 0 & 0 & 0 & V_1(s) \\
	V_4(s) & -V_3(s) & V_2(s) & -V_1(s) & 0 \\
\end{array}
\right)\,.
	\end{equation}
\end{widetext}
	The matrices satisfy $\mathcal{M}_1^T=-\mathcal{M}_1$ and $\mathcal{M}_2^T=-\mathcal{M}_2$. The corresponding solutions are,
	\be
	\begin{split}
			\mathcal{A}_1(s)&=\exp\left(2\m\mathcal{M}_1(1-s)\right)\mathcal{A}_1(s=1)\,, \\
			 \mathcal{A}_2(s)&=\exp\left(\a\int_s^1 ds'\mathcal{M}_2(s')\right)\mathcal{A}_2(s=1)\,.
	\end{split}
	\ee
	where $\a=2\m/(1+\m)$. Due to the properties of the matrices $\mathcal{M}_{1,2}$, we can write,
	\be
	\mathcal{A}_{1,2}(s)^T\mathcal{A}_{1,2}(s)=\mathcal{A}_{1,2}(s=1)^T\mathcal{A}_{1,2}(s=1)\,,
	\ee
	as constants evaluated at $s=1$. The metric of measure is independent of the affine parameter $s$ and it follows that the complexity is,
	\be
	C(U)=\sqrt{\mathcal{A}_1^T\mathcal{A}_1+(1+\m)\left(\mathcal{B}^T\mathcal{B}+\mathcal{A}_2^T\mathcal{A}_2\right)}\,.
	\ee
	
	\subsection{\texorpdfstring{$k=2$}{Lg}}
 
	For this case, the constants of motion form the subset,
	\be
	\mathcal{B}=\left(v_1\,,v_2\,,v_3\,,v_4\,,v_5\,,v_6\,,v_7\,,v_8\,,v_9\,,v_{10}\right)\,.
	\ee
	The remaining variables form the vector, 
	\be
	\mathcal{A}(s)=\left(V_{11}(s)\,,V_{12}(s)\,,V_{13}(s)\,,V_{14}(s)\,,V_{15}(s)\right)\,,
	\ee
	satisfying,
	\be
	\frac{d\mathcal{A}(s)}{ds}+\frac{2\m}{1+\m}\mathcal{M}\mathcal{A}(s)=0\,,
	\ee
	with the matrix,
	\be
	\mathcal{M}=\left(
	\begin{array}{ccccc}
		0 & -v_{10} & v_9 & -v_8 & -v_4 \\
		v_{10} & 0 & -v_7 & v_6 & v_3 \\
		-v_9 & v_7 & 0 & -v_5 & -v_2 \\
		v_8 & -v_6 & v_5 & 0 & v_1 \\
		v_4 & -v_3 & v_2 & -v_1 & 0 \\
	\end{array}
	\right)\,,
	\ee
	satisfying $\mathcal{M}^T=-\mathcal{M}$. The solution is given by,
	\be
	\mathcal{A}(s)=\exp\left(\a \mathcal{M}(1-s)\right)\mathcal{A}(s=1)\,,
	\ee
	with $\a=2\m/(1+\m)$. Again, the norm of the vector $\mathcal{A}(s)$ is independent of the affine parameter $s$, and the complexity,
	\be
	C(U)=\sqrt{\mathcal{B}^T\mathcal{B}+(1+\m)\mathcal{A}^T\mathcal{A}}\,.
	\ee
	
	\subsection{\texorpdfstring{$k=3$}{Lg}}
 
	Finally, for the three local cases, the only constant of motion is,
	\be
	\mathcal{B}=\left(v_{15}\right)\,.
	\ee
	while the remaining variables form the vector,
	\be
	\mathcal{A}(s)=\left(V_1(s)\,,V_2(s)\,,\dots\,,V_{14}(s)\right)\,,
	\ee
	which satisfies, 
	\be
	\mathcal{A}(s)=\exp\left(2\m\mathcal{M}(1-s)\right)\mathcal{A}(s=1)\,.
	\ee
	where,
	\be
	\mathcal{M}=\left(
	\begin{array}{cccccccc}
		0 & 0 & 0 & 0 & 0 & 0 & 0 & -v_{15} \\
		0 & 0 & 0 & 0 & 0 & 0 & v_{15} & 0 \\
		0 & 0 & 0 & 0 & 0 & -v_{15} & 0 & 0 \\
		0 & 0 & 0 & 0 & v_{15} & 0 & 0 & 0 \\
		0 & 0 & 0 & -v_{15} & 0 & 0 & 0 & 0 \\
		0 & 0 & v_{15} & 0 & 0 & 0 & 0 & 0 \\
		0 & -v_{15} & 0 & 0 & 0 & 0 & 0 & 0 \\
		v_{15} & 0 & 0 & 0 & 0 & 0 & 0 & 0 \\
	\end{array}
	\right)\,.
	\ee
	Fortunately, for the one-local case, the exponentiation can be done exactly. yielding the $(8\times8)$ dimensional ``magic matrix". In this case, the complexity takes the simple form,
	\be
	C(U)=\sqrt{\mathcal{A}^T\mathcal{A}+(1+\m)v_{15}^2}\,.
	\ee
	For the four-local case, $V_i(s)=v_i$ for all $i=1\dots 15$.
	
		\begin{figure}
		\centering
		\includegraphics[scale=0.50]{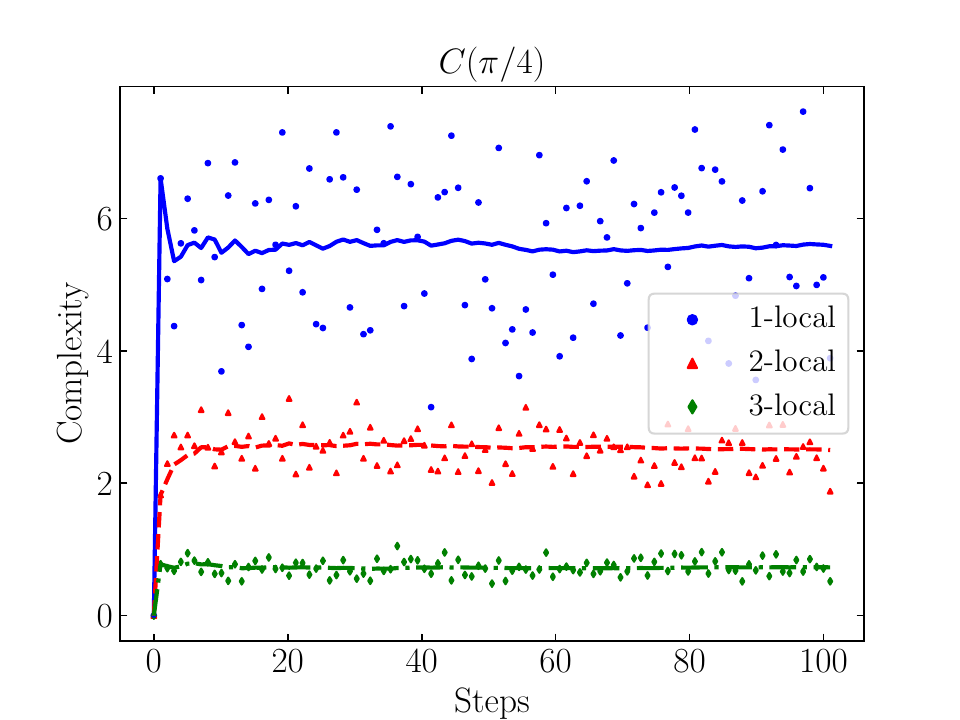}
		\caption{``Direct" complexity as function of unitary operator $U_\text{target}(t)$ for $\theta=\pi/4$ for $k=1,2,3$ local operators. The scattered colored points are the actual values derived, whereas the dotted colored lines denote the stepwise averaged values.}
		\label{comp_U}
	\end{figure}
	
	\subsection{Two definitions of complexity}\label{twodef}
 
	The notion of Nielsen complexity that we study here, is understood as a measure of the number of gates needed to construct the target state starting from a reference state. However, as it is defined, it cares about only the reference and the target state. For these two states, we have two boundary conditions, namely $U(s=0)=\mathcal{I}$ and $U(s=1)= U_\text{target}$, where $U_\text{target}$ takes the reference state to the target one. Below, we describe two separate ways to apply the machinery of Nielsen complexity. The two methods vary in the choices of the initial states at different steps.

	\subsubsection{Direct complexity}
	
	In our quantum walk, at some timestep $n$, our target state can be arbitrarily close to the reference state. What we mean is that the target unitary corresponding to a purified state at timestep $t=n$ can be very close to identity if identity is associated with the two-qubit state $|0\ra\otimes|0\ra$. Then, there is no way in this setup to guarantee that the circuit corresponding to this complexity also has other steps corresponding to the states associated with the walk for $m<n$ steps. In fact, Nielsen's complexity ensures that we find the complexity/circuit corresponding to the smallest possible circuit (geodesic in the space of the unitaries) connecting identity to the target unitary. Hence, if we always associate the identity to one particular reference state $|0\ra\otimes|0\ra$, this complexity never takes into consideration the previous states of the walk while finding the optimized circuit between the reference state and the state at some timestep $t=n$. This is what we call direct complexity. Here we only change the target state at each timestep, keeping the reference state unchanged. The complexities computed at each timestep, therefore, show a relative complexity in comparison to the chosen reference state. However, as explained above, it does not really care whether each of the optimized circuits at a certain timestep $t=n$ contains all the previous states along the walk.

			\begin{figure}
		\centering
		\includegraphics[width = 0.95\linewidth]{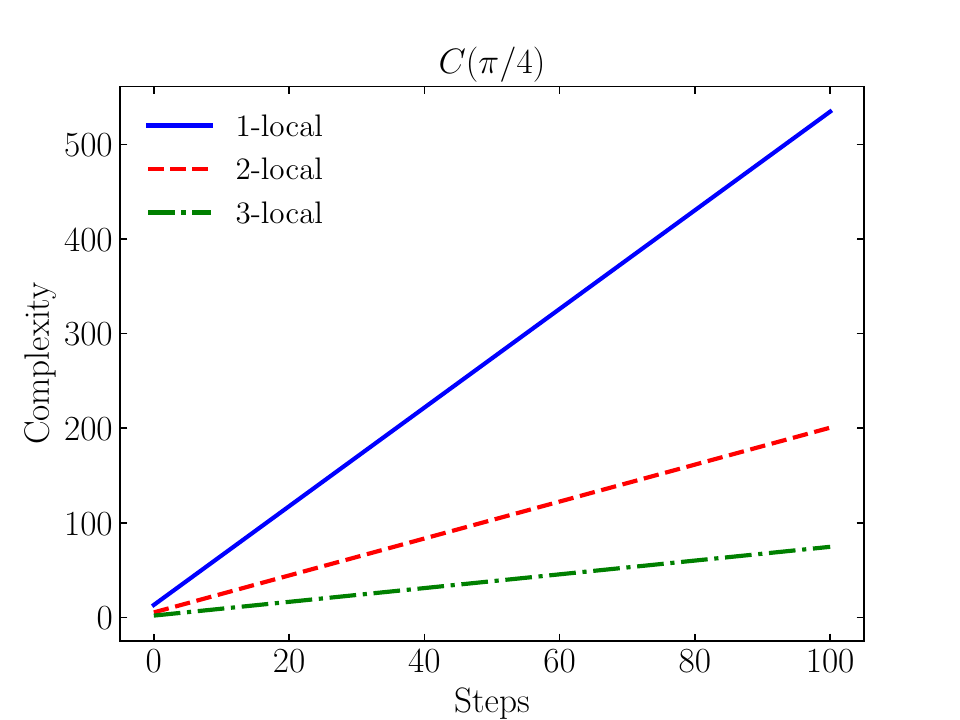}
		\caption{Complexity for ``stepwise" evolution for $\theta=\pi/4$ for $k=1,2,3$ local operators.}
		\label{comp_step}
	\end{figure}

	The corresponding plots for the $k=1,\, 2,\, 3$ local cases are shown in Fig.~\ref{comp_U}. We observe that the complexities for different timesteps with respect to the reference state behave in an uncorrelated, fluctuating way. However, it is worth noting that the fluctuating values decrease as we increase the notion of locality in the picture. In an explicit way, this basically means assigning fewer penalty factors to more and more generators of the $\text{SU}(4)$ group. From a gate perspective, this can be understood as more and more quantum gates becoming easily available as we increase the locality.

	\subsubsection{Stepwise complexity}\label{subsec: Stepwise complexity}

	From an explicit circuit construction perspective, it is more practical to consider complexities in a ``stepwise'' manner. In what follows, we explain what we mean by the term ``stepwise''. 
	
	Let us say that we first compute the complexity for the circuit, transforming the reference state to the purified state corresponding to the mixed state from the quantum walk at timestep $t=1$. Let us assume that the corresponding unitary is $U_1$, and the complexity computed turns out to be $C_1$. Now what we want to make sure of while considering the complexity of the purified state corresponding to the quantum walk at timestep $t=2$ is that we reach this state through the state at timestep $t=1$. This is more pragmatic from a circuit construction point of view in the sense that we want to simulate the full quantum walk through our circuit. To do this, we actually compute the complexity $C_2$ for the state at $t=2$ as $(C_1+C_2^{\prime})$, where $C_2^{\prime}$ is the complexity for the unitary $U_2^{\prime}$ connecting the states at $t=1$ and $t=2$. This unitary will not be the unitary $U_2$ connecting our actual reference state to the state at $t=2$. However once we find $U_2$, it is easy to find out $U_2^{\prime}$ since 
	\be
	U_2=U_2^{\prime} U_1 \implies U_2^{\prime}=U_2 U_1^{\dagger}.
	\ee
	Now, once we use the target unitary for the second step to be $ U_2^{\prime}$ instead of $U_2$, we find the complexity $C_2^{\prime}$ corresponding to the optimized circuit between the timesteps $t=1$ and $t=2$. 
	
	This step can be used for arbitrary steps in the same way. Once we have the unitaries $U_n$ connecting any $n^\text{th}$ timestep of the walk to our initial reference state, we can always find the unitary $U_n^{\prime}$connecting the unitary connecting the random-walk states at $t=(n-1)$ and $t=n$. 
	\be
	U_n^{\prime}= U_n U_{n-1}^{\dagger}.
	\ee
	The corresponding complexity for $U_n^{\prime}$ is then $C_n^{\prime}$ and we define the combined stepwise complexity for the $n$th step, ensuring that the circuit includes all previous random-walk states, as 
	\be
	C_n=C_1+C_2^{\prime}+C_3^{\prime}+\cdots C_n^{\prime}.
	\ee

	The plot corresponding to this is shown in Fig.~\ref{comp_step}. We find a linear growth of complexity with steps in this case. The growth persists forever, which is meaningful from a circuit construction point of view. This circuit successfully simulates all the states along the quantum walk at different steps. Here also, we find that the slope of the curves decreases as we increase the locality. It, therefore, seems to be universally true that as we make more and more generators (or equivalently quantum gates from a circuit perspective), it takes fewer number of gates to construct the optimized circuit.

	\begin{figure}
		\centering
		\includegraphics[width = 0.95\linewidth]{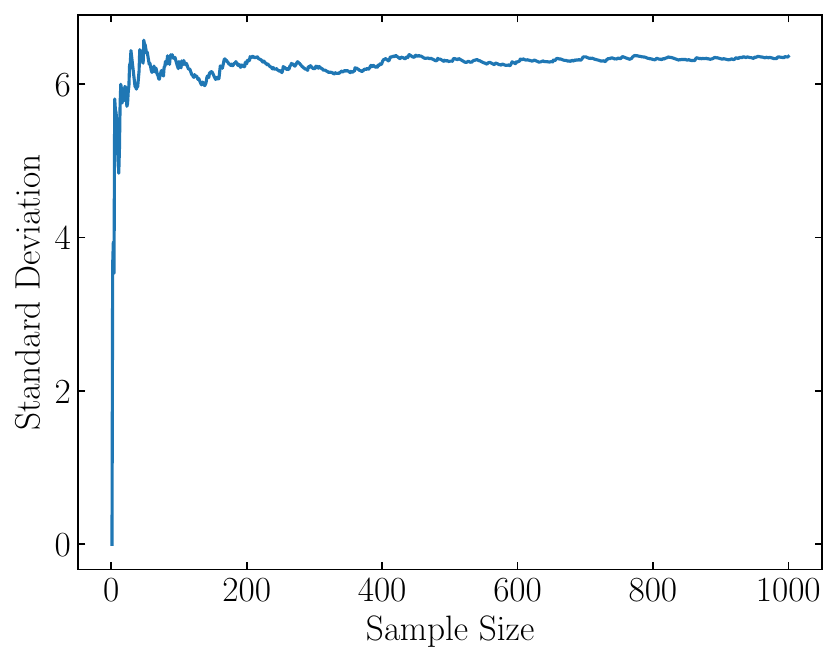}
		\caption{Standard deviation of the circuit complexity with an increasing sample size of target unitary operators calculated for the 10th step of DTQW.}
		\label{fig:standard_deviation}
	\end{figure}

	In  Fig.~\ref{fig:standard_deviation}, we plot the standard deviation of the complexity data with respect to the sample size we chose. This sample size refers to the number of random vector choices made for each timestep to generate the last three columns of the target unitary. Given a particular timestep (for example, $t=10$ in Fig.~\ref{fig:standard_deviation}), we vary the number of samples chosen to check the stability of our complexity. We find that around a sample size of $200$, the standard deviation stabilizes. Therefore, it is meaningful to choose a sample size of the order of $500$. Among all the $500$ samples for a given timestep $t=n$, and therefore $500$ unitaries, we choose the one for which the complexity is least. This is to make sure that we are making the most optimized choice out of the whole sample space of unitaries for each given timestep.

\begin{figure*}
		\centering
		\includegraphics[width = 0.95\linewidth]{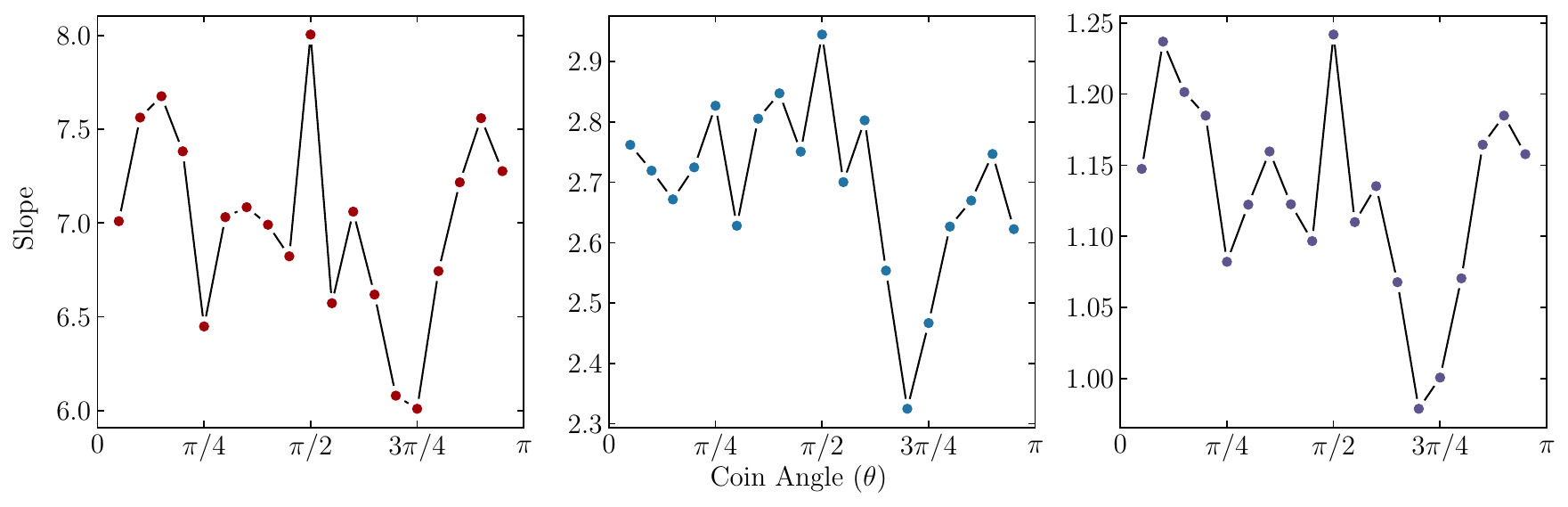}
		\caption{Slope of linear complexity with varying coin angle $\theta$ with \textbf{Left} : 1-local operators, \textbf{Middle} : 2-local operators and \textbf{Right} : 3-local operators. }
		\label{fig:Coin_angle_vs_slope}
	\end{figure*}

	In Fig.~\ref{fig:Coin_angle_vs_slope}, we plot the slope of the stepwise complexity plots for different locality notions with varying coin angles. Whereas the slope values increase with decreasing locality, which is expected from plots in Fig.~\ref{comp_step} already, we notice that there is a dip in the slope for each of the cases around coin angle value $2\pi/3$. These plots, therefore, indicate that among different coin angles, the complexity is least for the coin with angle $\theta \approx 2\pi/3$.

	\subsection{Fermionic Hamiltonian in continuum limit}
	Before concluding the section, we would like to point out that diagonalization of the two-particle fermionic Hamiltonian in \eqref{2phamiltonian}, leads to the following construction,
	\be
	V(s)=v_5 T_5+v_{10} T_{10}+v_{15} T_{15}=-i H_2(p_1,p_2)t
	\ee
	which solves for $v_5=0$ and,
	\begin{widetext}
	\begin{align}
	\begin{split}
		v_{10}&=-\frac{t}{2}\left(\sqrt{m^2+p_1^2+p_2^2-\sqrt{\left(m^2+2p_1^2\right)\left(m^2+2p_2^2\right)}}+\sqrt{m^2+p_1^2+p_2^2+\sqrt{\left(m^2+2p_1^2\right)\left(m^2+2p_2^2\right)}}\right)\,,\\
		v_{15}&=\frac{t}{2}\left(\sqrt{m^2+p_1^2+p_2^2-\sqrt{\left(m^2+2p_1^2\right)\left(m^2+2p_2^2\right)}}-\sqrt{m^2+p_1^2+p_2^2+\sqrt{\left(m^2+2p_1^2\right)\left(m^2+2p_2^2\right)}}\right)
	\end{split}
\end{align}
	\end{widetext}
	for $m=\sin\theta$. Consequently,
	\be
	C=\int^{\Lambda_1}\int^{\Lambda_2} dp_1 dp_2 \sqrt{v_{10}^2+v_{15}^2}\simeq \frac{5}{24}~t~\Lambda^3\ln\Lambda  
	\ee
	This complexity grows linearly with time. However, this complexity does not indicate the quantum walk completely but only an approximation in the continuum limit. Hence the complexity does not demonstrate the nuances of the walk completely. 
 
	\section{Quantum circuit}\label{circuit}

	\begin{figure}
		\centering
		\includegraphics[width = 0.7\linewidth]{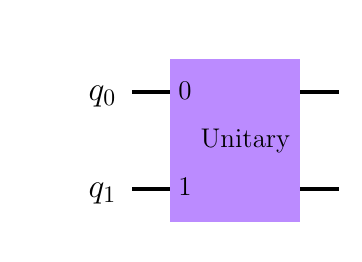}
		\caption{A general two-qubit quantum circuit for the step-wise unitary operator.}
		\label{q_circ}
	\end{figure}

 In this section, we connect the linear growth of cumulative step-wise complexity found in Sec.~\ref{subsec: Stepwise complexity} to constant circuit depth of explicit quantum circuit associated with target unitary. The target unitary operator can be associated with a two-qubit circuit, as shown in Fig.~\ref{q_circ}. To find the circuit depth of a quantum circuit, one is required to decompose the unitary into a universal set of gates. For our purpose, we will consider a one-qubit gate $U_3(\theta,\phi,\lambda)$ and two-qubit CNOT gate with the explicit forms as following, 

	\be
	U_3(\theta,\phi,\lambda)=\left(\begin{array}{cc} \cos\theta & e^{i\phi}\sin\theta\\ -e^{-i\phi}\sin\theta & e^{-i(\phi+\l)}\cos\theta\end{array}\right)\,,
	\ee
	and,
	\be
	\text{CNOT}=\left(\begin{array}{cccc}1&0&0&0\\0&0&0&1\\0&0&1&0\\0&1&0&0\end{array}\right)\,.
	\ee

	\begin{figure*}
		\centering
		\includegraphics[scale = 0.75]{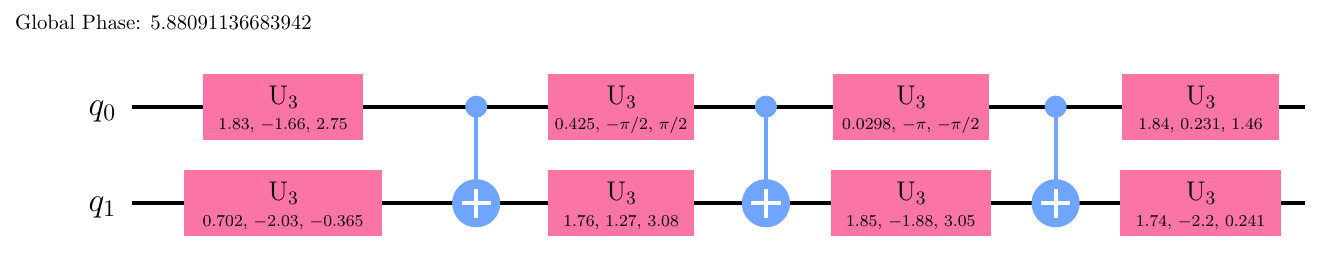}
		\caption{Representing a two-qubit quantum circuit for the step wise unitary operator using \texttt{Qiskit}.}
		\label{q1_circ}
	\end{figure*}
 
\begin{figure}
		\centering
		\includegraphics[scale = 0.50]{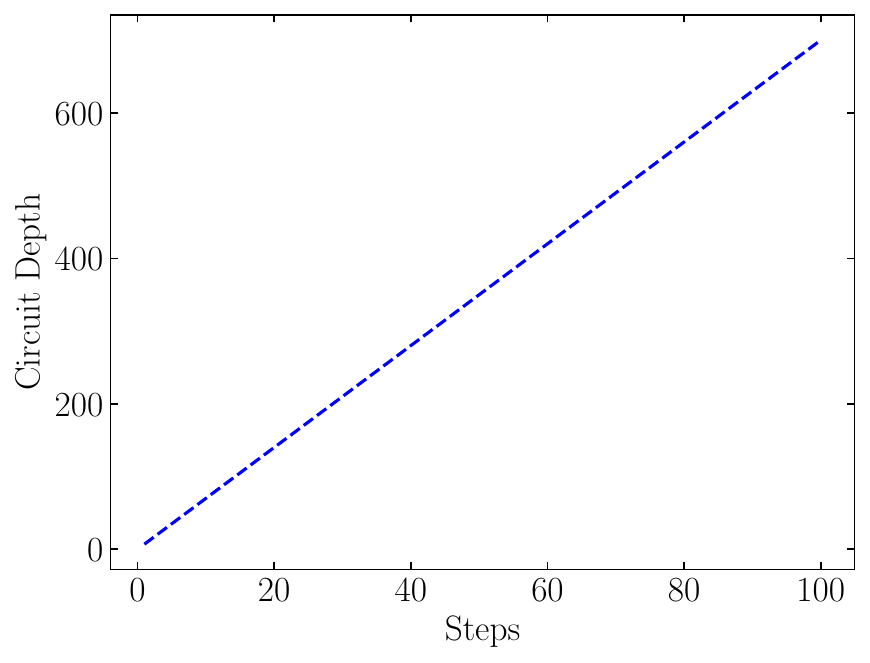}
		\caption{Quantum circuit depth corresponding to target unitary operator corresponding to 2-local case (estimated with explicit construction of quantum circuit using \texttt{Qiskit}) with varying time-steps of DTQW. }
		\label{fig:CircuitDepth_Steps}
	\end{figure}

   In Fig.~\ref{q1_circ}, we showed the explicit circuit associated with the target unitary for a particular walk step constructed using \texttt{Qiskit}\,\cite{Qiskit}. The circuit associated with the target unitary for different walk steps has a contact depth of $7$ layers with parameters $\theta,\phi,$ and $\lambda$ changing values.
    Therefore, similar to the direct complexity study using Nielsen's proposal, the cost of constructing the unitary seems almost a constant function. Therefore, if we again construct the circuit stepwise and cumulatively sum the depth of the individual circuits, the depth grows linearly again with the steps (see Fig.~\ref{fig:CircuitDepth_Steps}), in agreement with the complexity computed from Nielsen's proposal. 

It's important to emphasize that relation between $k$-local case in context of Nielsen's complexity (NC) and $k$-qubit gate in context of quantum circuit is not very well understood. If we naively consider that both are equivalent to each other, then $k=1$ should correspond to single qubit quantum circuits for target unitary. However, since the universal set of quantum gates atleast requires a two-qubit gate\,\cite{nielsen_chuang_2010} as known from the Solovay-Kitaev theorem, one can not construct a general $n$-qubit quantum circuit ($n>1$) with just single qubit gates (unless target unitary is separable into $n$ independent single qubit gates which is not true generally or in our case). The case $k=3$ on the other hand can be realised for $n$ qubit quantum circuits with $n\geq3$. However, in our case, we have a two-qubit target unitary, which can not be written in terms of three qubit quantum gates. These facts as mentioned above suggest that $k$-local operators should not in general be understood as $k$-qubit gates, and their exact relation needs further investigation. However, for the two-qubit circuit, our result indeed shows a qualitative similarity between the two concepts. It will be interesting to extend these studies to higher qubit circuits, where one can explicitly check if such similarities exist for $k\geq 3$.

	\section{Discussions}\label{discussions}
 
	We conclude the work with a brief account of what has been answered and what more remains to be done. To begin with:
	\begin{itemize}
		\item We have computed the complexity for the one-dimensional quantum walk using a $\text{SU}(2)$ coin. The walk entangles the position and internal degrees of freedom and produces a mixed state on partial tracing over the position degrees of freedom. Consequently to measure the complexity of the mixed state, we first canonically purify the mixed state and then evaluate the complexity using an approximate two-qubit quantum circuit.
		\item We compute and compare the complexities of the purified state using both the direct evolution operator and the step-wise evolution in the quantum walk. The complexity function oscillates with the steps around a mean value which can be associated with the depth of an average quantum circuit. The step-wise evolution, however, connects more with the actual quantum circuit and the quantum walk picture since the direct evolution ignores the steps in connecting the geodesic from the initial to the final step. As such, the step-wise evolution is a direct implementation of time-ordering and a successful simulation of the DTQW using a quantum circuit. The complexity of the step-wise evolution cumulatively grows with the steps and is indicative of the growing size of an associated quantum circuit and its complexity. 
		\item To give some context, we also implement a schematic quantum circuit using one and two qubit quantum gates to implement the step-wise unitary evolution. The circuit has constant depth and relates to the average complexity in Fig.~\ref{comp_U}. 
		
		\item Another upshot of doing the stepwise evolution is that although we have stepwise geodesics instead of a full one, it can produce for us the stepwise response functions which are valid for individual timesteps. Let us say if we want to write the Hamiltonian acting between steps $(n-1)$ and $n$, we can simply pick the corresponding stepwise unitary $U_n^{\prime}$ and get an estimate of the Hamiltonian as 
		\be
		H(n)=-\frac{1}{i}\ln [U_n^{\prime}]\, ,
		\ee
		since each timestep is of length one. Now we can write this Hamiltonian as \be
		H(n)=\sum_i V_i(n)T_i
		\ee
		to figure out which generator was effective and how much during a particular timestep. Finally, we can sum all those stepwise Hamiltonians with the corresponding step functions and write down a complete Hamiltonian that is time-independent stepwise but gives rise to all the purified states corresponding to the mixed one of the actual DTQW. This is somewhat analogous to finding out the response functions for different quantum gates in the Nielsen picture of complexity. This is a trivial task and the states being individually randomly distributed, these functions do not show up any particular nature of growth or decay. However, it might produce further interesting results in the case of an explicitly chaotic quantum walk\,\cite{Omanakuttan:2021aa} or for a time-dependent coin operator\,\cite{Sen:2023zyx}.
	\end{itemize}
	However, more questions have been uncovered by the exploration. Some of the pressing questions, which we could not answer in this work due to lack of resources, but intend to complete them in immediate future follow-up works are:
	\begin{itemize}
		\item First of all, the precise connection between the circuit picture and the continuum formulation is still largely opaque. We have just implemented an example circuit that can connect with the step-wise evolution. However, the exact nature of how the geodesic length is connected with the actual quantum circuit still remains to be explored further.
		\item The distinction between the step-wise evolution and the unitary evolution is based on the logic that one can view the step-wise evolution with some quantum circuits and hence the size of the circuit grows along with its complexity which exhibits itself in the linear growth. However, for real quantum systems, the complexity grows linearly with time for early time and smooths out to a constant. One reason for the discrepancy might be the fact that the dimension of the Hilbert space of the state in quantum walk linearly grows with time. In order to gain insights into this apparent conflict, one way to move forward would be to connect the quantum walk to the Hamiltonian of some physical system, to get a more realizable connection with real-time systems. Consequently designing quantum circuits for the quantum walk will act as a bridge to gain more insights into the mapping of field theory complexity with the actual circuit compiling complexity. 
		
	\end{itemize}

	\section{Acknowledgements}
	The work of AB is supported by the Polish National Science Centre (NCN) grant 2021/42/E/ST2/00234. KS is partially supported by FAPESP grant 2021/02304-3.

	\appendix
	
	\section{Details of generators}\label{generators}
	We provide the explicit form of the fermionic generators $\g_a$ here. There are 

	\begin{align}
		\begin{split}
			\g_1&=\s_1\otimes I_2=\left(
			\begin{array}{cccc}
				0 & 0 & 1 & 0 \\
				0 & 0 & 0 & 1 \\
				1 & 0 & 0 & 0 \\
				0 & 1 & 0 & 0 \\
			\end{array}
			\right)\,,\\
			  \g_2 &=\s_2\otimes I_2=\left(
			\begin{array}{cccc}
				0 & 0 & -i & 0 \\
				0 & 0 & 0 & -i \\
				i & 0 & 0 & 0 \\
				0 & i & 0 & 0 \\
			\end{array}
			\right)\,,\\ 
			\g_3&=\s_3\otimes\s_1=\left(
			\begin{array}{cccc}
				0 & 1 & 0 & 0 \\
				1 & 0 & 0 & 0 \\
				0 & 0 & 0 & -1 \\
				0 & 0 & -1 & 0 \\
			\end{array}
			\right)\,, \\
   \g_4&=\s_3\otimes\s_2=\left(
			\begin{array}{cccc}
				0 & -i & 0 & 0 \\
				i & 0 & 0 & 0 \\
				0 & 0 & 0 & i \\
				0 & 0 & -i & 0 \\
			\end{array}
			\right)\,.
		\end{split}
	\end{align}

	The $\g_a$ satisfy,
	\be
	\{\g_a,\g_b\}=2\eta_{ab}\,.
	\ee
	The explicit forms of the generators $T_i$ for the $\text{SU}(4)$ group are then given by,

	\begin{equation}
		\begin{split}
			T_i&=\{\g_1\,,\g_2\,,\g_3\,,\g_4\,,i\g_1\g_2\,,i\g_1\g_3\,,i\g_1\g_4\,,i\g_2\g_3\,, \\
			& \qquad  i\g_2\g_4\,,i\g_3\g_4\,,-i\g_1\g_2\g_3\,,-i\g_1\g_2\g_4\,, \\
			& \qquad \qquad -i\g_1\g_3\g_4\,,-i\g_2\g_3\g_4\,, -\g_1\g_2\g_3\g_4 \}
		\end{split}
	\end{equation}

\vfill

\bibliographystyle{apsrev4-1}
\bibliography{ref}

\end{document}